\documentclass[runningheads]{llncs}
\usepackage[T1]{fontenc}
\usepackage{graphicx}
\usepackage{cite}
\usepackage{amsmath,amssymb,amsfonts}
\usepackage{algorithmic}
\usepackage{graphicx}
\usepackage{textcomp}
\usepackage{xcolor}
\usepackage{multicol}
\usepackage{multirow}
\usepackage{makecell} 
\usepackage{booktabs} 
\usepackage{array}
\usepackage{caption}
\usepackage{wrapfig}

\usepackage{graphicx}
\usepackage{subcaption}

\usepackage{float}  

\usepackage{tikz}
\usetikzlibrary{arrows.meta, positioning}

\usepackage{url}

\usepackage[linesnumbered,ruled,vlined]{algorithm2e}

\usepackage{xspace}
\usepackage{booktabs}
\newcommand{\tools}[1]{\textsc{#1}}
\newcommand{\fastLEC}{\tools{FastLEC}\xspace}
\newcommand{\dac}{\tools{PDP-CEC}\xspace}
\newcommand{\hybridcec}{\tools{HybridCEC}\xspace}
\newcommand{\phcec}{\tools{pHCEC}\xspace}
\newcommand{\acec}{\tools{ABC \&cec}\xspace}
\newcommand{\sprove}{\tools{\&splitprove}\xspace}

\begin{document}
\title{FastLEC: Parallel Datapath Equivalence Checking with Hybrid Engines}
%
%

\author{Xindi Zhang\inst{1,2}\and
Furong Ye\inst{3}\and
Zhihan Chen\inst{1,2}\and
Shaowei Cai\inst{1,2}}
\authorrunning{X. Zhang et al.}
%
\institute{
Key Laboratory of System Software (Chinese Academy of Sciences) and State Key Laboratory of Computer Science, Institute of Software, Chinese Academy of Science \and School of Computer Science and Technology, University of Chinese Academy of
Sciences, Beijing, China \email{\{zhangxd,chenzh,caisw\}@ios.ac.cn} \and
LIACS, Leiden University \email{f.ye@liacs.leidenuniv.nl}}

\maketitle
\begin{abstract}

Combinational equivalence checking (CEC) remains a challenge EDA task in the formal verification of datapath circuits due to their complex arithmetic structures and the limited capability or scalability of SAT, BDD, and exact-simulation (ES) based techniques when used independently. This work presents \fastLEC, a hybrid prover that unifies these three formal reasoning engines and introduces three strategies that substantially enhance verification efficiency. First, a regression-based engine-scheduling heuristic predicts solver effectiveness, enabling more accurate and balanced allocation of computational resources. Second, datapath-structure-aware partitioning strategies, along with a dynamic divide-and-conquer SAT prover, exploit the regularity of arithmetic designs while preserving completeness. Third, the memory overhead of ES is significantly reduced through address-reference-count tracking, and simulation is further accelerated through a GPU-enabled backend. \fastLEC is evaluated across 368 datapath circuits. Using 32 CPU cores, it proves 5.07× more circuits than the widely used \acec tool. Compared with the latest best datapath-oriented serial and parallel CEC provers, \fastLEC outperforms them by 3.33× and 2.67× in PAR-2 time, demonstrating an improvement of 74 newly solved circuits. With the addition of a single GPU, it achieves a further 4.07× improvement. The prover also demonstrates excellent scalability.

\end{abstract}

\section{Introduction\label{sec:pre}}

Combinational Equivalence Checking (CEC) is a fundamental formal verification technique used to determine whether two combinational netlists compute the same Boolean function. It plays a critical role in modern Electronic Design Automation (EDA) flows and is widely employed during logic synthesis, optimization, and circuit formal verification and testing~\cite{vasicek2017relaxed,jarratt2011engineering,mishchenko2008scalable}. The predominant industrial paradigm is the SAT-sweeping framework; tools such as the \&cec engine in Berkeley ABC~\cite{mishchenko2007abc} are based on this methodology.

Despite their success on control-oriented logic, SAT-sweeping techniques exhibit inherent limitations on arithmetic-intensive datapath circuits~\cite{seger2021formal}. Structured arithmetic components, e.g., compressor trees, Booth encoders, partial-product networks, create dense XOR dependencies~\cite{chowdhary1999extraction,long2013lec}, while synthesis-induced divergence~\cite{kamaraju2010power} further disrupts structural matching~\cite{kaufmann2020incremental}.
Formally, high XOR depth and long carry chains suppress CDCL propagation, yielding branching-sensitive and unstable SAT runtimes. Conversely, BDDs suffer exponential blow-up when wide adder or multiplier cones enlarge the PI-cut, causing decision-diagram widths to exceed memory despite dynamic reordering. Datapath CEC thus remains a class of instances where neither SAT nor BDD provides robust worst-case behavior.



Several recent efforts have proposed enhanced reasoning engines for datapath CEC. Exact simulation (ES) complements SAT by efficiently handling difficult node pairs, enabling hybrid frameworks such as \hybridcec. Parallelization has been explored through CPU/GPU collaboration~\cite{chatterjee2010equipe}, partitioning~\cite{possani2019parallel}, while more recent approaches such as \phcec~\cite{chen2025datapath} and \dac~\cite{zhou25dac} extend HybridCEC using parallel reasoning engines or metis-based~\cite{karypis1997metis} graph partitioning. Although these techniques offer substantial improvements, many equivalence pairs in large datapath circuits remain hard to prove, and the scalability of existing methods is still insufficient for increasingly complex industrial designs.

At the core of our approach is the observation that no single reasoning paradigm, whether SAT, BDD, or exact simulation (ES), scales uniformly across all datapath structures. Each engine exhibits complementary strengths that can be exploited if orchestrated intelligently.

Building on this insight, we introduce \fastLEC, a unified hybrid CEC framework that integrates SAT solving, BDD reasoning, and ES into a coordinated verification workflow. The framework leverages datapath-aware structural information, adaptive engine scheduling, and resource-conscious simulation strategies to systematically reduce the complexity of hard-to-prove equivalence pairs while preserving soundness and completeness.

Our contributions are as follows:

\begin{enumerate}
    \item \textbf{A hybrid verification framework with predictive engine scheduling.}
    We integrate SAT solving, exact simulation (ES), and binary decision diagram (BDD) reasoning into a single coordinated CEC framework.
    A regression-based predictor estimates engine effectiveness on demand, enabling informed and balanced scheduling among heterogeneous solvers and significantly improving resource efficiency.

    \item \textbf{A memory-efficient, GPU-accelerated exact-simulation backend.}
    We reduce ES memory consumption through address-reference-count tracking and further accelerate simulation using a GPU-enabled backend, yielding substantial throughput improvements for arithmetic-heavy designs.

    \item \textbf{A datapath-aware partitioning strategy combined with a parallel divide-and-conquer SAT engine.}
    By exploiting the structural regularity of arithmetic datapath circuits, we design partitioning techniques tailored to arithmetic topologies and couple them with a parallel SAT-based decomposition procedure that preserves completeness while reducing the complexity of hard equivalence pairs.

\end{enumerate}

We use a benchmark suite of 368 datapath miters to evaluate \fastLEC against the widely used CEC prover \acec~\cite{mishchenko2007abc}, the leading datapath-oriented solver \hybridcec~\cite{chen2024datapath}, and a recent parallel datapath prover \dac~\cite{zhou25dac}.
Under a 32-core configuration, \fastLEC solves 74 more instances than the best existing competitor. In terms of PAR-2 performance, it achieves speedups of 7.73$\times$, 3.33$\times$, and 2.67$\times$ over \acec, \hybridcec, and \dac, respectively.

\fastLEC also exhibits substantial scalability. The 128-thread configuration verifies 44.88\% more instances than the sequential version. Notably, the 32-thread solver attains speedups of 100$\times$, 500$\times$, and 1000$\times$ on \textbf{33}, \textbf{19}, and \textbf{14} miters, respectively.
Furthermore, augmenting the 32-thread version with a single NVIDIA~4090 GPU provides an additional 4.07$\times$ speedup.


\vspace{-0.35cm}

\section{Preliminaries}

\subsection{Combinational Equivalence Checking}

\textit{Combinational Equivalence Checking} (CEC) is the task of determining whether two combinational circuits compute the same Boolean function.

\textit{Datapath circuits}, prevalent in microprocessor design, contain dense arithmetic logic (e.g., multipliers, adders, and multiplexers) with high XOR density, making equivalence checking particularly difficult~\cite{lv2012formal,chen2023integrating}.
In this work, circuits are represented as latch-free \textit{And-Inverter Graphs} (AIGs), where each primary input (PI) and gate output is mapped to a Boolean variable. 
For ease of algorithm design and evaluation, this paper converts the AIG into a more compact XOR-AND (inverter) Graphs (XAGs) representation.
The \textit{transitive fan-in cone} (TFI) of a node $g$ includes all nodes that have a path leading to $g$.

To check equivalence, a \textit{miter} circuit is built by XOR-ing corresponding POs of the two circuits and OR-ing the results; the circuits are equivalent iff the miter output is identically zero for all inputs.

\vspace{-0.35cm}
\subsection{Proof Engines}
Several compelete methods have been employed to verify the output constant-zero property for given miters.
Boolean Satisfiability (SAT) solvers~\cite{kuehlmann2004dynamic}, Binary Decision Diagram (BDD) compilers~\cite{kuehlmann2002robust}, Exact Simulation (ES) provers~\cite{wu2008novel}, and their hybrid methods~\cite{kuehlmann2002robust,chen2023integrating} are among the most commonly used approaches.

\vspace{-0.25cm}
\paragraph{\textbf{SAT Solving}}

Given Boolean variables \( V = \{v_1, \ldots, v_n\} \), a \textit{literal} is a variable or its negation, a \textit{clause} is a disjunction (\( \vee \)) of literals, and a \textit{Conjunctive Normal Form (CNF)} formula is a conjunction (\( \wedge \)) of clauses. The satisfiability (SAT) problem asks whether there exists an assignment to \( V \) that satisfies the CNF.

AIGs and XAGs can be translated into CNF via Tseitin transformation~\cite{tseitin1983complexity}, with miter checking encoded by adding a clause asserting the output is zero.

Modern SAT solvers are based on the conflict-driven clause learning (CDCL) framework, a systematic search algorithm equipped with powerful reasoning techniques. 
Parallel solving typically follows two paradigms: clause-sharing portfolios, which run diversified solvers concurrently~\cite{zhang2022parkissat,schreiber2024mallobsat}, and partitioning-based methods like cube-and-conquer~\cite{heule2011cube}, which split the CNF into many subproblems solved in parallel—particularly effective for hard combinatorial instances~\cite{heule2016solving}.


\vspace{-0.25cm}
\paragraph{\textbf{Exact Simulation} (ES)}

It is a method for enumerating all the possible assignments of PI and checking the corresponding output of the miter.

A straightforward approach is to explicitly enumerate all Boolean assignments (0/1) for each primary input (PI). 
Meanwhile, more efficient methods exploit SIMD-style propagtion or probability-based simulation~\cite{agrawal1996characteristic,wu2006potential}.

Meanwhile, alternative methods based on floating-point number propagation exploit the principles of Single Instruction, Multiple Data (SIMD) to accelerate the evaluation process. 
The runtime or memory usage of an ES engine proposed by \cite{wu2006potential} is exponentially related to the number of PIs $N$ and linearly related to the number of gates. 

\vspace{-0.25cm}
\paragraph{\textbf{Binary Decision Diagram} (BDD)}
The miter's output function can be encoded as a BDD. If the compiled BDD reduces to the constant-zero terminal, then the two circuits are equivalent. Otherwise, any non-zero path corresponds to a distinguishing input pattern. BDD-based checking is exact and efficient for small or structured circuits, but suffers from exponential growth on datapath logic~\cite{kuehlmann2002robust}. 
CUDD~\cite{somenzi1998cudd} and Sylvan~\cite{van2015sylvan} are representative sequential and parallel BDD packages, respectively.

\vspace{-0.35cm}
\subsection{Standard Sweeping Flow}


\begin{wrapfigure}{r}{0.5\textwidth} 
    \vspace{-10pt} 
    \centering
    \includegraphics[width=0.98\linewidth]{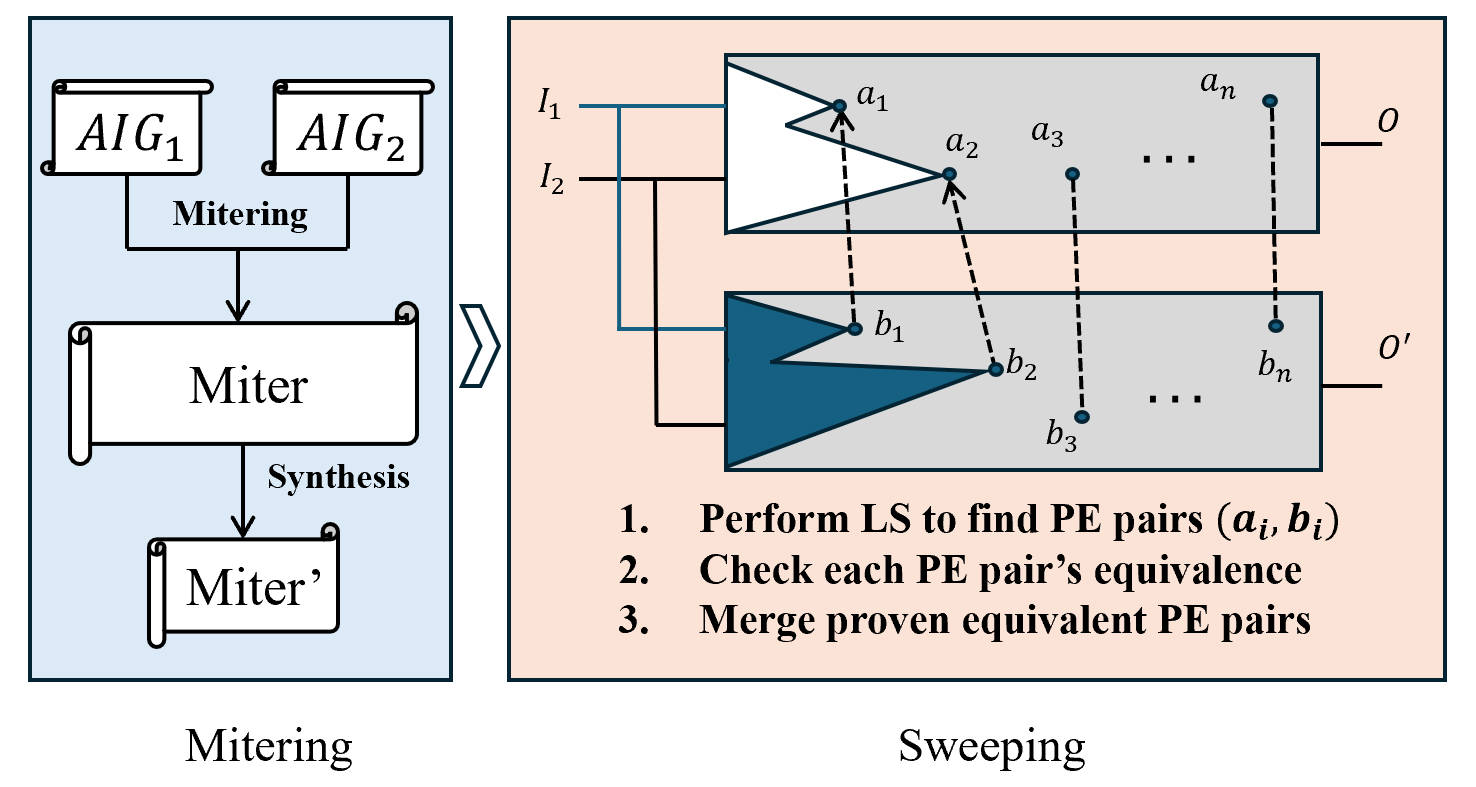}
    \caption{Overviewing Sweeping Flow.}
    \label{fig:sweeping}
\end{wrapfigure}

A typical sweeping flow for CEC is presented in Fig.~\ref{fig:sweeping}.
To check equivalence, a miter is constructed and simplified using logic synthesis~\cite{bjesse2004dag,kuehlmann2002robust,mishchenko2005fraigs}, then refined via a sweeping flow~\cite{kuehlmann2002robust,kuehlmann2004dynamic}. The process includes: (1) logic simulation to identify potentially equivalent (PE) node pairs; (2) localized checking of PE pairs using SAT, BDD, or ES; and (3) circuit simplification by merging proven pairs. 

Recent work~\cite{chen2023integrating} shows that exact simulation (ES) excels on small-input, XOR-heavy cones, complementing SAT. An XOR-density score is used to guide engine selection, improving performance on datapath circuits. This heuristic remains effective in recent parallel datapath CEC solvers~\cite{chen2024datapath,zhou25dac}.

\vspace{-0.35cm}
\subsection{Regression Models and Hyperparameter Optimization}

The Gradient-boosted regression trees (GBRT) framework has been widely used in\textit{supervised learning} due to its efficiency in capturing non-linear dependencies and supporting heterogeneous features.
Its well-known and optimized implementation in \textsc{XGBoost}~\cite{chen2016xgboost} has been widely adopted across various applications and has been a top contender in Kaggle competitions. This competitive performance and its mechanism for handling heterogeneous features make XGBoost well-suited for predicting solver performance on structured inputs.

While the performance of machine learning models, including XGBoost, is known to be sensitive to their hyperparameter settings, various hyperparameter optimization techniques have been proposed.
Among them, SMAC3~\cite{JMLR:v23:21-0888}, which is based on Bayesian Optimization, has emerged as a standard tool for tuning machine learning models and optimizers such as SAT solvers.

\section{Hybrid Scheduling–Based CEC Framework}

\subsection{Motivation: Absence of a Dominant Reasoning Engine}

SAT is effective on large synthesized logic but becomes unstable on datapath circuits, where arithmetic regularities and dense XOR interactions often cause severe slowdowns, as evidenced in Table~\ref{tab:mult-runtime}. BDDs perform well on small unsynthesized datapath cones but experience abrupt blow-ups once structural complexity increases due to synthesis. ES exhibits far more stable behavior and strong parallel scalability. Table~\ref{tab:hardcases} shows that ES can solve many instances that defeat both SAT and BDD with the aid of CPU or GPU parallelism; the techniques will be discussed later. 

These complementary yet non-overlapping strengths give rise to the \texttt{No-Gold Selection Phenomenon}: no single engine dominates across all datapath structures, motivating an adaptive and instance-aware hybrid scheduling framework.
\begin{table}[t]
\centering
\begin{minipage}{0.48\textwidth}
\centering
\caption{Runtime (s) of SAT, ES, and BDD on multiplier miters.}\label{tab:mult-runtime}
\setlength{\tabcolsep}{2pt}
\scalebox{0.6}{
\begin{tabular}{c|c|cccc|c|c}
\toprule
Miter & SAT & \multicolumn{4}{c|}{CPU-ES} & GPU-ES & BDD \\
      &     & 1T & 32T & 64T & 128T &       &     \\
\midrule
8*8   & 0.45     & $<0.01$ & $<0.01$ & $<0.01$ & $<0.01$ & $<0.01$ & $<0.01$ \\
9*9   & 2.40     & $<0.01$ & $<0.01$ & $<0.01$ & $<0.01$ & 0.01    & $<0.01$ \\
10*10 & 22.79    & $<0.01$ & $<0.01$ & $<0.01$ & $<0.01$ & 0.01    & $<0.01$ \\
11*11 & 85.75    & $<0.01$ & $<0.01$ & $<0.01$ & $<0.01$ & 0.01    & 0.01 \\
12*12 & 434.88   & 0.10    & $<0.01$ & $<0.01$ & $<0.01$ & 0.01    & 0.66 \\
13*13 & NA       & 1.94    & $<0.01$ & $<0.01$ & $<0.01$ & 0.01    & 2.70 \\
14*14 & NA       & 10.23   & 0.17    & $<0.01$ & $<0.01$ & 0.05    & 7.53 \\
15*15 & NA       & 52.23   & 1.58    & 0.88    & 0.37    & 0.29    & 25.11 \\
16*16 & NA       & 229.37  & 6.58    & 4.36    & 2.41    & 2.06    & 85.80 \\
17*17 & NA       & 1029.06 & 34.42   & 18.00   & 9.63    & 10.57   & 257.90 \\
18*18 & NA       & NA      & 139.64  & 77.75   & 39.07   & 46.93   & 783.67 \\
19*19 & NA       & NA      & 690.54  & 347.43  & 174.53  & 215.89  & NA \\
20*20 & NA       & NA      & 2779.61 & 1535.40 & 790.08  & 983.99  & NA \\
\bottomrule
\end{tabular}
}
\end{minipage}\hfill
\begin{minipage}{0.48\textwidth}
\centering
\caption{Runtime comparison on representative hard cases.}\label{tab:hardcases}
\setlength{\tabcolsep}{8pt}
\scalebox{0.65}{
\begin{tabular}{c|c|c|c|c}
\toprule
Miter & Instance & SAT & GPU-ES & BDD \\
\midrule
11*11 & 17\_29\_9  & 27.58  & \textbf{0.01} & 1.24 \\
11*11 & 14\_22\_8  & 35.47  & \textbf{0.01} & NA \\
12*12 & 21\_19\_10 & 135.27 & \textbf{0.01} & 10.13 \\
12*12 & 17\_17\_0  & NA     & \textbf{0.01} & NA \\
13*13 & 13\_25\_25 & 121.53 & \textbf{0.01} & 261.73 \\
13*13 & 15\_14\_13 & 883.17 & \textbf{0.01} & NA \\
14*14 & 29\_18\_2  & NA     & \textbf{0.06} & NA \\
15*15 & 17\_18\_17 & NA     & \textbf{0.28} & NA \\
16*16 & 17\_29\_4  & NA     & \textbf{1.98} & NA \\
16*16 & 15\_20\_4  & NA     & \textbf{2.03} & NA \\
17*17 & 17\_32\_53 & NA     & \textbf{9.61} & NA \\
18*18 & 17\_12\_7  & NA     & \textbf{41.54} & NA \\
\bottomrule
\end{tabular}
}
\end{minipage}
\end{table}

\vspace{-1cm}
\subsection{The Hybrid Scheduling Framework}

\begin{figure}[t]
\centerline{\includegraphics[width=1.0\textwidth]{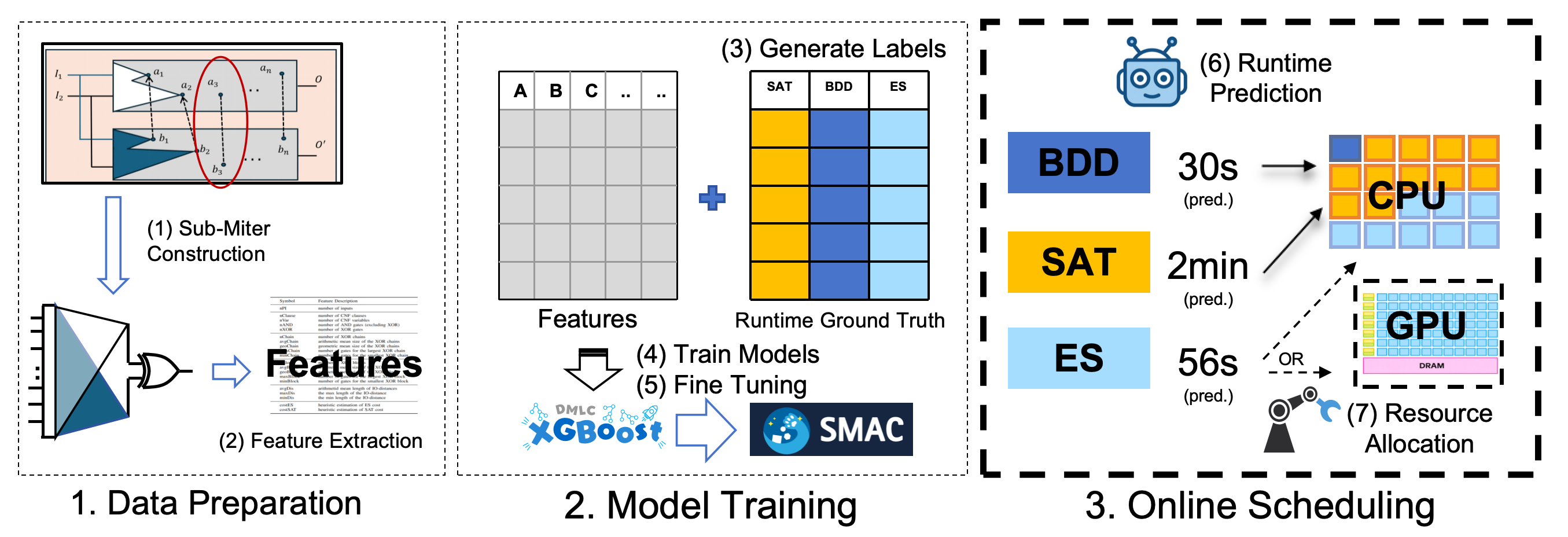}}
\caption{The Scheduling Framework of FastLEC. \label{fig:framework}}
\label{fig}
\end{figure}

\vspace{-0.15cm}
\paragraph{\textbf{Overview.}}
Fig.~\ref{fig:framework} overviews the organization of our 
hybrid-engine scheduling method.  
The key objective is to exploit the complementary algorithmic strengths 
of three verification back-ends, SAT, ES, and BDD, while avoiding the pathological behaviors that each engine can exhibit on specific circuit structures.
To achieve this, our approach is divided into three stages:
\emph{data preparation}, \emph{offline model training}, and 
\emph{online scheduling}.  
The offline stages construct a learned cost model
that predicts the runtime of each engine on a given sub-miter; 
the online stage uses these predictions to drive a resource-aware 
scheduling policy on heterogeneous compute hardware.

\vspace{-0.25cm}
\paragraph{\textbf{Data Features Preparation.}}

Given a miter circuit, we generate a set of \emph{sub-miters} by virtually simulating a sweeping pass. Each target signal pair is isolated by retaining only its fanin cones and aggressively merging topologically preceding nodes, thereby encoding localized equivalence obligations and producing compact subcircuits comparable to those obtained through standard sweeping.

For every sub-miter, we extract a compact feature vector describing its
structural, logical, and arithmetic characteristics.
The features include:

(i)~Structural Indicators such as XAG features (No. 1--4), CNF features (No. 5--7), distances to any input, the output, and their sum(No. 16-22), and fanout degrees (No. 23--24);

(ii)~Datapath descriptors capturing XORs (No. 8--15). 
Two XOR gates are in the same \textit{XOR block} if they are directly connected; 
In this paper, two gates are in the same \textit{XOR chain }if they are connected by an XOR gate; 

(iii)~Cost Estimations for SAT and ES, the same method as \cite{chen2023integrating} (No. 25--26). 
\texttt{cost\_ES} =  $2^{\texttt{num\_PIs}};$
\texttt{cost\_SAT} = $\sum_{b \in \texttt{XOR\_Blocks}} 2^{|b|}$;

(iv)~Logic Simulation-based Characteristics (No. 27--32). 
\texttt{Stability} measures the signal bias of each node from random simulation, where signals whose values remain consistent across all simulation rounds are considered maximally stable, while \texttt{entropy} quantifies information uncertainty using Shannon entropy (higher values indicate greater uncertainty).

These features collectively form a heterogeneous representation that 
captures the behaviors most relevant to the three verification engines.

\begin{table}[htbp]
\centering
\caption{Feature List for XAG Circuit Analysis (Total: 32 features)}
\label{tab:features}
\scriptsize
\scalebox{0.8}{
\begin{tabular}{m{1.5cm} m{9cm}}
\toprule
\textbf{No.} & \textbf{Features} \\
\midrule
1--4 & \texttt{num\_PIs}, \texttt{num\_gates}, \texttt{num\_XOR\_gates}, \texttt{num\_AND\_gates} (basic XAG statistics) \\
\midrule
5--7 & \texttt{num\_CNF\_vars}, \texttt{num\_CNF\_clauses}, \texttt{num\_CNF\_lits} (CNF characteristics) \\
\midrule
8--11 & \texttt{min/max/avg/geo\_mean\_XOR\_block} (XOR block statistics) \\
\midrule
12--15 & \texttt{min/max/avg/geo\_mean\_XOR\_chain} (XOR chain statistics) \\
\midrule
16--22 & \texttt{max/avg\_idis} (distance to inputs), \texttt{max/avg\_odis} (distance to outputs), \texttt{max/min/avg\_sum\_dis} (combined distance) \\
\midrule
23--24 & \texttt{max/avg\_out\_degree} (out-degree statistics) \\
\midrule
25--26 & \texttt{cost\_SAT}, \texttt{cost\_ES} (solver costs estimate heuristic as in \cite{chen2023integrating}) \\
\midrule
27--29 & \texttt{min/max/avg\_stability} (stability of simulation) \\
\midrule
30--32 & \texttt{min/max/avg\_entropy} (entropy during simulation) \\
\bottomrule
\end{tabular}
}
\end{table}

\vspace{-0.25cm}
\paragraph{\textbf{Model Training.}}
To obtain supervised learning targets, we execute all three engines on the 
training set under controlled time (20 minutes) and record their 
actual runtimes. 
This produces a ground-truth runtime matrix indexed by 
(sub-miter, engine), forming a dataset of triples 
$(\mathsf{features}, \mathsf{engine}, \mathsf{runtime})$.

We use XGBoost~\cite{chen2016xgboost} to train regression models that predict the one-core CPU time of SAT and BDD on a given sub-miter. The models are trained on $7{,}892$ sub-miters, obtained by sampling 10\% from the original $179{,}478$ PE pairs and discarding instances with $\texttt{num\_PIs} \le 22$. Each sub-miter is represented by 32 extracted features.


Using 10-fold cross-validation with each fold trained on 80\% of the dataset and tested on the remaining 20\%, our predictors, i.e., regression models, achieved Mean Squared Errors (MSEs)\footnote{MSE is the average squared error between predictions and true values; both ground-truth and predicted CPU times are capped at $1{,}200$s.} of $1{,}462$ for SAT and $49{,}684$ for BDD, which are largely inflated by capped runtimes. Restricting to samples with SAT/BDD CPU times below $1{,}200$s can reduce the MSEs to $463$ and $14{,}550$. The pronounced disparity indicates that SAT runtimes are substantially more predictable from the given features, likely reflecting its more stable behavior across instances. Note that all reported results are obtained with the predictors in which hyperparameters were fine-tuned by SMAC3~\cite{JMLR:v23:21-0888} with a budget of 500 trials (Table~\ref{tab:xgb-para}). This hyperparameter optimization process yielded a $10\%$ ($1{,}601$ to $1{,}462$) MSE improvement for SAT and $25\%$ ($62{,}202$ to $49{,}684$) MSE improvement for BDD.

\begin{table}[t]

\caption{XGBoost Parameters.\label{tab:xgb-para}}
\begin{center}

\setlength{\tabcolsep}{3pt}
\scalebox{0.8}{
\begin{tabular}{|l|c|c|c|c|c|c|c|}
\hline
Parameters & learning rate & max depth & subsample & colsample & lambda & alpha & number of trees \\
\hline
Domains 
& (0.01, 0.3)
& (3, 10)
& (0.5, 1.0)
& (0.5, 1.0)
& (0, 10.0)
& (0.0, 5.0)
& (50, 500)
\\
\hline
\end{tabular}
}

\end{center}
\end{table}


\vspace{-0.25cm}
\paragraph{\textbf{Runtime Prediction and Scheduling Method}}
During verification, each sub-miter is featurized and evaluated by the trained models to estimate its one-core costs for SAT and BDD, denoted as $\hat{t}_{\text{SAT}}$ and $\hat{t}_{\text{BDD}}$, respectively.

Unlike the NP-hard and highly variable SAT/BDD runtimes, ES runtime is deterministically governed by circuit size and input space. It depends linearly on the number of gates and exponentially on the number of primary inputs, and is thus readily predictable. We model it as
\(
\hat{t}_{\mathsf{ES}} = \alpha \cdot \texttt{num\_gates} \cdot 2^{(\texttt{num\_PI} - \beta)},
\)
with coefficients fitted from the training data.\footnote{In the Sect. 4.1 setting, the best-fit parameters are $\alpha=0.0003$ and $\beta=23$.}

These runtime predictions are consumed by a scheduling module that distributes the $n$ available cores across engines according to the following principles:
\begin{itemize}
    \item \textit{Engine feasibility.} 
    SAT is always enabled when $n \ge 2$; 
    Enable ES when $\hat{t}_{\mathsf{ES}}/n \leq \gamma \cdot \mathsf{cutoff}$, where $\gamma = 1.5$ \footnote{Safety marjins in this paper are tuned in [0.5--2]$\times$ which is empirically robust, since variations of this scale fall within normal prediction noise and therefore do not materially affect performance while preventing mis-allocation.} serves as a safety margin against prediction error;
    BDD is enabled only when $\hat{t}_{\mathsf{BDD}} < \varphi \cdot \hat{t}_{\mathsf{SAT}}$, where the safety margin is $\varphi = 0.8$ and at least two threads are available.

    \item \textit{CPU scheduling.}
    The BDD engine, when enabled, receives at most one thread due to its substantial memory overhead, while still offering useful complementarity on certain large, unsynthesized circuits.  
    To determine the SAT–ES split, we evaluate the ratio
    \(
    \rho = {\hat{t}_{\mathsf{SAT}}}/{n \cdot \hat{t}_{\mathsf{ES}}}.
    \)
\footnote{
We use the $n$-thread ES cost as it reflects the expected parallel runtime; when predictions are noisy and speedups non-linear, enabling additional engines yields risk without guaranteed benefit. In contrast, our parallel SAT scheme preserves the root task on its original thread, ensuring that parallelization rarely degrades the single-thread baseline if ignoring memory and system overheads.
}
    Ideally, all resources should be devoted to the engine with the smallest predicted cost; however, due to inherent prediction noise, we introduce a margin that treats runtimes within a bounded range as equivalent.
    
    (i) When $\rho < 1/2$, SAT is predicted to be more promising; thus ES and BDD each receive one thread, and all remaining threads are assigned to SAT;
    
    (ii) When $\rho \in [1/2,\, 2]$, the engines are potentially complementary, so threads are divided evenly between SAT and ES, with at most one SAT thread reallocated to BDD;
    
    (iii) When $\rho > 2$, ES is predicted to be more promising; thus SAT and BDD each receive one thread, and all remaining threads are assigned to ES.

    \item \textit{GPU-accelerated scheduling.}
    GPU-based ES is modeled as 128 CPU threads and enabled when its predicted time exceeds 0.1 s. When enabled, ES is offloaded to the GPU, one CPU thread is assigned to BDD, and the remaining threads to SAT; otherwise, the scheduler reverts to the CPU-only allocation.

    \item \textit{Special-case handling.}
    For easy instances $\hat{t}_{\mathsf{ES}}/n \leq 0.1 s$, only ES is enabled. When $n=1$, only SAT and ES are enabled with the selection heuristic according to $score_{XOR} = cost_{SAT}/cost_{ES}$, same as \cite{chen2023integrating}.

\end{itemize}

By integrating this hybrid engine selection method, a CEC prover named \fastLEC has been developed.

\vspace{-0.25cm}
\paragraph{\textbf{Scheduling Never Degrades Correctness.} } The scheduling policy affects only how engines are invoked, not the semantics of the proof procedures. SAT, BDD, and ES remain individually sound and complete, and scheduling neither alters the miter nor modifies any inference rule. In particular, the parallel SAT engine always retains the unsplit root CNF on its original thread, ensuring that all partitioned tasks are conservative refinements. Thus, mispredictions may impact runtime but cannot compromise soundness or completeness.

\vspace{-0.25cm}
\subsection{GPU-Accelerated Exact Simulation Engine}

\begin{figure*}[t]
  \centering
  \begin{minipage}[t]{0.45\textwidth}
    \centering
    \includegraphics[width=\linewidth]{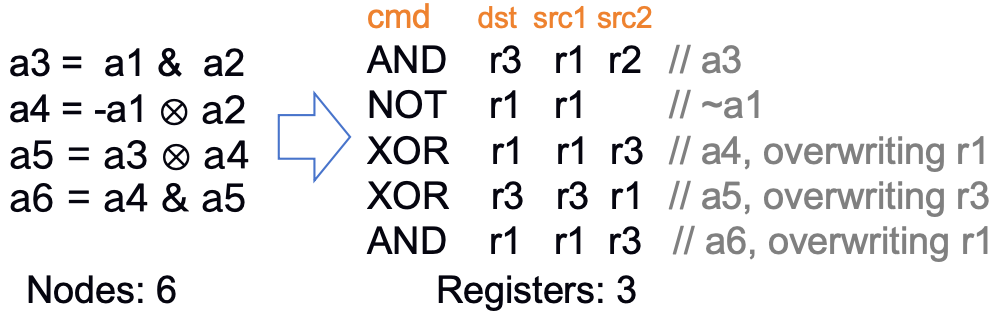}
    \caption{Instructionalized Example.}
    \label{fig:instr}
  \end{minipage}\hfill
  \begin{minipage}[t]{0.5\textwidth}
    \centering
    \vspace{-2cm}
    \captionof{table}{Compression achieved by instructionalization on multiplier miters.}
    \label{tab:compression}
    \scalebox{0.6}{
    \begin{tabular}{|l|r|r|r|}
      \hline
      \textbf{\makecell{Multiplier \\ Miter}} &
      \textbf{\makecell{Vars \\(Before)}} &
      \textbf{\makecell{Instrs \\(After)}} &
      \textbf{\makecell{Compression \\Ratio}} \\
      \hline
      8$\times$8  & 1463  & 61  & 4.16\% \\
      \hline
      12$\times$12 & 3647  & 99  & 2.71\% \\
      \hline
      16$\times$16 & 6822  & 143 & 2.09\% \\
      \hline
      24$\times$24 & 16178 & 234 & 1.44\% \\
      \hline
      32$\times$32 & 29553 & 309 & 1.04\% \\
      \hline
    \end{tabular}
    }
  \end{minipage}
\end{figure*}

\vspace{-0.15cm}
\paragraph{Instructionalized ES. }
To make exhaustive simulation feasible on GPUs, the circuit is compiled into a compact, branch-free instruction stream, with registers aggressively reused under reference-count guidance. This transformation reduces memory footprint to the minimal cut of the circuit rather than its full structural size, enabling GPU execution even for large datapath miters. As illustrated in Fig.~\ref{fig:instr}, reference-count guided register reuse can shrink a six-node circuit to only three registers. This compression is essential because the key limitation of GPU execution is the extremely small amount of on-chip memory available per SM. For larger datapath circuits, the effect becomes even stronger. As reported in Table~\ref{tab:compression}, the number of instructions grows much more slowly than the total gate count, since it is governed by the minimal cut of the circuit rather than by its full structural size.

During simulation, the instruction stream is stored in the constant cache of each SM and all bit-vector registers of a thread reside in shared memory. The kernel then executes a simple sequence of bitwise operations without branching or pointer lookup. 
CPU-based ES typically propagates $2^{14}$ input vectors in parallel, as suggested by our preliminary experiments, whereas, in this paper, each GPU thread processes only 32 due to its limited per-thread on-chip memory.
Through this combination of structural compression and highly regular parallel execution, exhaustive evaluation becomes efficient even for large multiplier miters that would otherwise overwhelm GPU resources.




\vspace{-0.35cm}
\paragraph{Effectiveness.}
The combination of structural compression and branch-free SIMD execution yields orders-of-magnitude acceleration on large multipliers. As shown in Table~\ref{tab:mult-runtime}, a single GPU delivers throughput comparable to roughly one hundred CPU cores. Nevertheless, profiling reveals that the dominant performance bottleneck remains global-memory traffic, indicating that further speedups hinge primarily on reducing memory reads and writes rather than on additional computation parallelism.




\vspace{-0.55cm}
\subsection{Structure-Aware Parallel SAT Engine}

\vspace{-0.25cm}

For circuits with a relatively large number of primary inputs, ES-based and BDD-based methods become ineffective, while Boolean SAT solving remains the most viable approach. Therefore, enhancing SAT techniques specifically designed for datapath circuits is of vital importance.

This subsection presents a parallel divide-and-conquer framework, equipped with a structure-aware partitioning heuristic, for efficient equivalence checking of datapath circuits.

\vspace{-1.25cm}
\subsubsection{Dynamic SAT Partitioning Framework.}

\begin{figure}[t]
  \centering
  \begin{minipage}[t]{0.44\textwidth}
    \centering
    \includegraphics[width=1.05\linewidth]{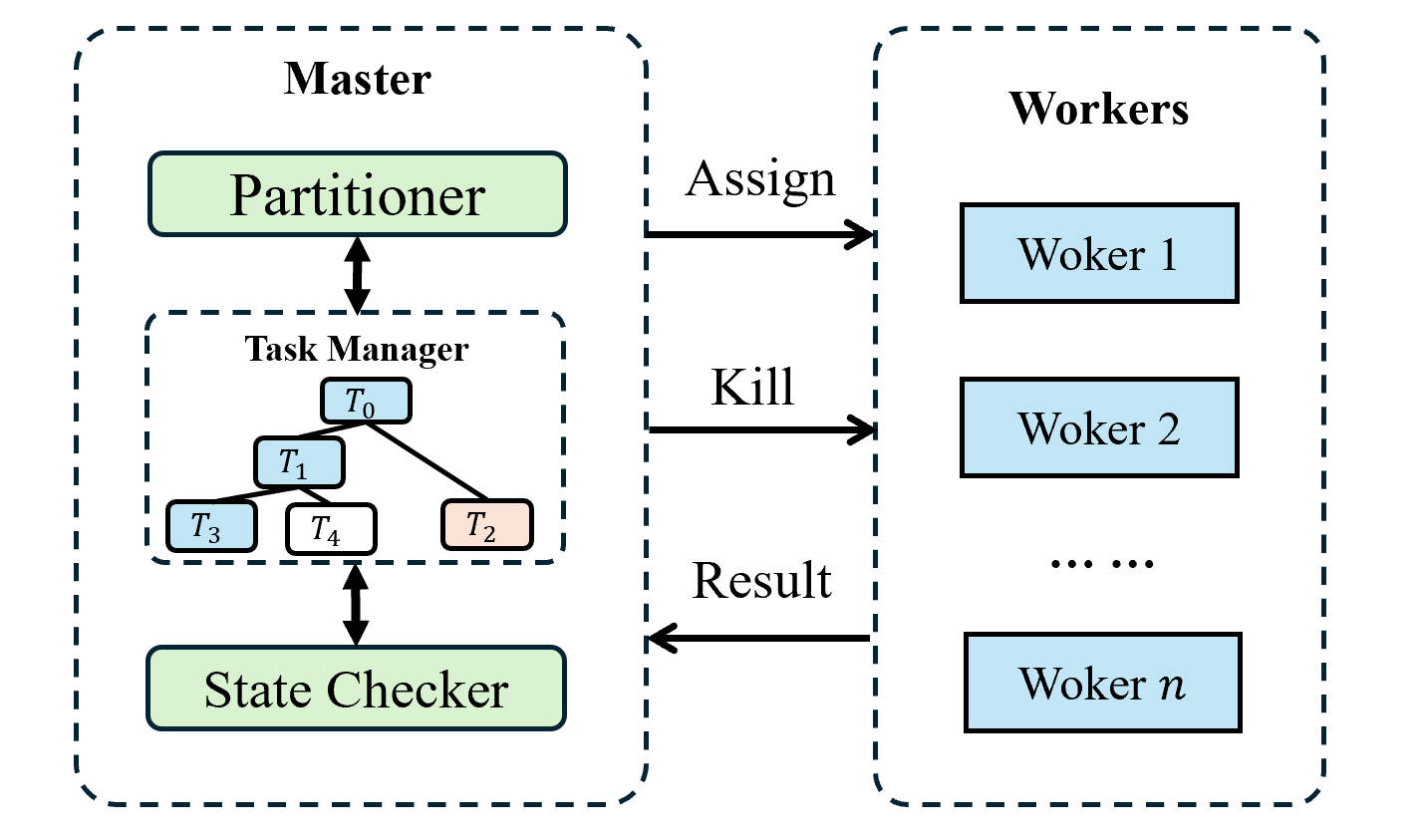}
    \caption{Parallel SAT Framework for Structure-Aware Partitioning}
    \label{fig:parallel}
  \end{minipage}\hfill%
  \begin{minipage}[t]{0.55\textwidth}
    \scriptsize
    \vspace{-3.3cm}
    \scalebox{0.95}{
    \renewcommand{\algorithmcfname}{\scriptsize Alg.}
    \begin{algorithm}[H]
        \caption{\scriptsize Structural Scoring for Partition}\label{alg:score}
        \KwIn{XAG netlist $N$, a cube $L$}
        \KwOut{Variables scores $sc(v)$}
            $N' \leftarrow \texttt{propagate\_and\_preprocess}(N, L)$\;
            $chains \leftarrow \texttt{extract\_XOR\_Chains}(N')$\;
            $\texttt{dis}_{I}, \texttt{dis}_{O} \leftarrow \texttt{compute\_distance}(N')$\;
            \For{$c$ in \texttt{XOR\_chains}}{
                $S \leftarrow \texttt{compute\_cut\_points}(N', c)$\;
                \For {$v$ in $c$}{
                    $d \leftarrow \alpha \cdot \texttt{dis}_{O}[v] + (1 - \alpha) \cdot \texttt{dis}_{I}[v] + 1.0$\;
                    $sc[v] \leftarrow |c|^2 / d$\;
                    \lIf{$v$ in $S$}{ $bump(v)$}
                }
            }
            $sc \leftarrow \texttt{score\_propagte}(sc, factor, round)$\;
            \Return sc\;
    \end{algorithm}
    }
  \end{minipage}
\end{figure}

As illustrated in Fig.~\ref{fig:parallel},
\fastLEC adopts a classical \emph{master--worker} execution model~\cite{DBLP:conf/cav/ZhaoCQ24}. 
Worker threads carry out sub-tasks and report their results, whereas the master handles
task generation, dependency tracking, and global state deduction.

\vspace{-0.15cm}
\begin{itemize}

\item \textit{Partitioner.}
    Selects an unsplit, long-running, and unsolved sub-task, chooses a splitting variable, and generates two child tasks, which are further simplified through Boolean propagation.

\item \textit{Task Manager.}
    Maintains the task hierarchy (dependencies, liveness, scheduling) and dispatches ready
    tasks to idle workers, ensuring that no worker receives a task that has already been solved.

\item \textit{State Checker.}
    A \texttt{SAT} result immediately certifies non-equivalence and halts the procedure.
    A sub-task proven \texttt{UNSAT} invalidates all its descendants; if all siblings are
    \texttt{UNSAT}, the parent is also marked \texttt{UNSAT}, enabling upstream propagation
    and early pruning of the task tree.




\end{itemize}






\vspace{-0.35cm}
\subsubsection{Structural Aware Boolean Partitioning}


\vspace{-0.5cm}
\paragraph{Observations and Explanations.} 

\fastLEC provides a toolkit for analyzing partitioning variables, measuring their actual
acceleration effect under specific task states and supporting visual interpretation.
Preliminary experiments show that, in XOR-dominated regions, variables nearer the primary outputs are
often effective partition points, as they naturally decompose the region into more balanced smaller
components. 

Since strict XOR-block detection is overly rigid, we approximate such
structures using XOR chains, their structural cuts, and the minimum distances to one PI and the
PO (\texttt{dis}$_I$, \texttt{dis}$_O$).

Long XOR chains dominate datapath dependencies; fixing a variable near the chain’s output cuts the largest portion of this structure. Distance terms approximate this reduction, so high-scoring variables yield the greatest collapse of XOR depth and, therefore, the most effective SAT decomposition.

\vspace{-0.25cm}
\paragraph{Partition Heuristics. } 
The task to be split is chosen as the unsolved leaf with the longest runtime, prioritizing
computational bottlenecks; each task is represented by a cube (a conjunction of literals).
The scoring algorithm for selecting the splitting variable is shown in
Algorithm~\ref{alg:score}. We first extract XOR gates and construct an XOR--AND Graph
(XAG), whose node indices can be aligned with CNF variables after redundant intermediate
nodes are removed.

Algorithm~\ref{alg:score} simplifies the netlist (line~1), extracts XOR chains (line~2),
and computes input/output distances (line~3). For each chain $c$, it identifies structural
cut points and increases their scores (lines~5 and~9). Each node receives a score
proportional to $|c|^2$ and inversely scaled by its topological distance (line~8), with a
bias toward nodes nearer the outputs ($\alpha = 0.6$). Consequently, longer XOR chains
contribute more prominently to the scoring.

We refine the variable scores through a propagation mechanism that diffuses scores across the circuit structure. Each variable's score is updated by averaging with its neighboring variables' scores, weighted by a self-influence factor. This process is abstracted in line 10, where $\text{factor}$ controls the self-weight and $\text{round}$ specifies the number of propagation iterations.

\vspace{-0.45cm}
\section{Experimental Evaluation}
\vspace{-0.15cm}

This section evaluates \fastLEC against state-of-the-art CEC provers, examines the effectiveness of its core techniques, and provides further analysis. 

\fastLEC is \textit{open-sourced} at \url{https://github.com/dezhangxd/fastLEC}.

\vspace{-0.35cm}
\paragraph{Settings and Competitors}
All experiments run on a cluster with 1 TB RAM, two AMD EPYC 9754 CPUs (128 physical cores), and one NVIDIA GeForce RTX 4090 (24G) graphics card under Ubuntu 20.04; parallel provers use 32 CPU cores by default. \fastLEC is implemented in C++17/C11, relies on XGBoost, AIGER, and KISSAT, and is compiled with g++ 9.4.0 using -O3. All competing CEC provers are written in C/C++ and compiled in the same environment.

The competitors being considered are:

\vspace{-0.25cm}
\begin{itemize}
\item \textbf{\acec}: A sequential SAT-sweeping CEC prover in Berkeley-ABC~\cite{mishchenko2007abc}, targeting general benchmarks.
\item \textbf{\hybridcec}~\cite{chen2023integrating}: A strong datapath-oriented prover that integrates exact simulation with SAT sweeping, effective on XOR-dense circuits.
\item \textbf{\sprove}: A parallel divide-and-conquer CEC tool built in ABC~\cite{mishchenko2007abc}.
\item \textbf{\phcec}~\cite{chen2024datapath}: Parallelizing \hybridcec by replacing its SAT solver with PRS~\cite{zhang2022parkissat} and parallelizing ES, often achieving super-linear speedups.
\item \textbf{\dac}~\cite{zhou25dac}: A parallel datapath prover inspired by \hybridcec, using graph partitioning and solving many cases beyond \hybridcec.
\end{itemize}

\vspace{-0.7cm}
\paragraph{Benchmarks}

The benchmarks used in this paper originate from the latest datapath EC track of the Integrated Circuit EDA Elite Challenge\footnote{\url{https://eda.icisc.cn/}}.
The suite consists of \underline{368} single-output equivalence miters derived from real-world datapath designs, and the full collection is available in the \fastLEC repository.
It represents a superset of the benchmarks used in prior studies, such as \cite{chen2023integrating, zhou25dac}, as provided by the competition organizers.
For evaluation, the benchmarks are divided into three difficulty classes: easy (127), medium (120), and hard (121).

\vspace{-0.35cm}
\paragraph{Metrics}

The wall-clock timeout for each prover is set to \texttt{1 hour}. To evaluate overall performance, we use the penalized average runtime (\texttt{PAR2} (in sec.)), a standard metric in formal methods competitions, where each unsolved miter is assigned twice the cutoff time, i.e., 2 hours.



    




\vspace{-0.35cm}
\subsection{Comparasions with State-of-the-arts}

\begin{table}[t]
\caption{Comparing \fastLEC with Top Sequential and Parallel CEC Provers. \label{tbl:sota}}
\begin{center}

\setlength{\tabcolsep}{9pt}
\scalebox{0.55}{
\renewcommand{\arraystretch}{0.8}

\begin{tabular}{crrrrcrrrr}
\toprule

Class &
Prover  &
\#Proved  &
PAR2&
Mono  &
Class &
Prover  &
\#Proved  &
PAR2&
Mono  \\
\midrule

\multirow{7}{*}{\makecell{Easy\\(127)}} 
& \fastLEC(32t+GPU)              & 127        & 8.66       & 0          &
\multirow{7}{*}{\makecell{Med-\\ium\\(120)}}  
& \fastLEC(32t+GPU)                  & 120        & 11.88      & 32                \\ \cmidrule(r){2-5} \cmidrule(r){7-10}

& \fastLEC(32t)                  & \textbf{127}        & 14.4       & 0          &
& \fastLEC(32t)                  & \textbf{120}        & \textbf{62.11}      & \textbf{32}         \\ \cmidrule(r){2-5} \cmidrule(r){7-10}
& \dac(32t)                      & 126        & 71.26      & 0          &
& \dac(32t)                      & 88         & 1930.22    & 0          \\
& \phcec(32t)                    & \textbf{127}        & \textbf{6.15 }      & 0          &
& \phcec(32t)                    & 87         & 1981.85    & 0          \\
& \sprove(32t)                   & 23         & 5897.08    & 0          &
& \sprove(32t)                   & 0          & 7200.0     & 0          \\ \cmidrule(r){2-5} \cmidrule(r){7-10}
& \hybridcec                     & 125        & 115.61     & 0          &
& \hybridcec                     & 88         & 2259.1     & 0          \\
& \acec                          & 67         & 3412.27    & 0          &
& \acec                          & 0          & 7200.0     & 0          \\

\midrule

\multirow{7}{*}{\makecell{Hard\\(121)}} 
& \fastLEC(32t+GPU)              & 121        & 401.68     & 69         &
\multirow{7}{*}{\makecell{All\\(368)}}  
& \fastLEC(32t+GPU)                & 368        & 138.94     & 101         \\ \cmidrule(r){2-5} \cmidrule(r){7-10}

& \fastLEC(32t)                  & \textbf{93}         & \textbf{2241.89}    & \textbf{41}         &
& \fastLEC(32t)                  & \textbf{340}        & \textbf{762.36 }    & \textbf{73}         \\ \cmidrule(r){2-5} \cmidrule(r){7-10}
& \dac(32t)                      & 52         & 4203.61    & 0          &
& \dac(32t)                      & 266        & 2036.18    & 0          \\
& \phcec(32t)                    & 52         & 4120.75    & 0          &
& \phcec(32t)                    & 266        & 2003.29    & 0          \\
& \sprove(32t)                   & 0          & 7200.0     & 0          &
& \sprove(32t)                   & 23         & 6750.35    & 0          \\ \cmidrule(r){2-5} \cmidrule(r){7-10}
& \hybridcec                     & 38         & 5368.01    & 0          &
& \hybridcec                     & 251        & 2541.58    & 0          \\
& \acec                          & 0          & 7200.0     & 0          &
& \acec                          & 67         & 5892.82    & 0          \\

\bottomrule
\end{tabular}
}
\end{center}
\end{table}






Table~\ref{tbl:sota} compares \fastLEC with two sequential CEC provers (\acec and \hybridcec) and three parallel provers (\sprove, \dac, and the parallel configuration of \phcec). 

All parallel tools are evaluated with 32 threads, and a suffix ($n$\texttt{t} or GPU) indicates the number of CPU threads or the use of a GPU. For each solver, we report the number of solved instances (\texttt{\#Proved}), the PAR2 score (PAR2), and the number of miters solved monolithically (\texttt{Mono}).

The results in Table~\ref{tbl:sota} demonstrate the strong performance of \fastLEC. Relative to the best sequential prover, \hybridcec, our solver achieves substantial improvements: \fastLEC is at least 5×, 10×, and 32× faster on 97, 70, and 33 instances, respectively.

\fastLEC also surpasses the latest parallel datapath-oriented CEC prover \dac presented at DAC'25. Overall, \fastLEC solves 74 more instances (27.82\%) than \dac, and on the most challenging hard-miter category, it verifies 41 additional miters (78.85\%), highlighting its superiority on structurally difficult arithmetic circuits.

For relatively easy instances, the runtime may appear slightly longer. This overhead arises from feature extraction, model prediction, and latency introduced by the engine timeout callback mechanism.

\begin{figure}[tbp]
    \centering
    \begin{minipage}{0.48\textwidth}
        \centering
        \includegraphics[width=0.55\linewidth]{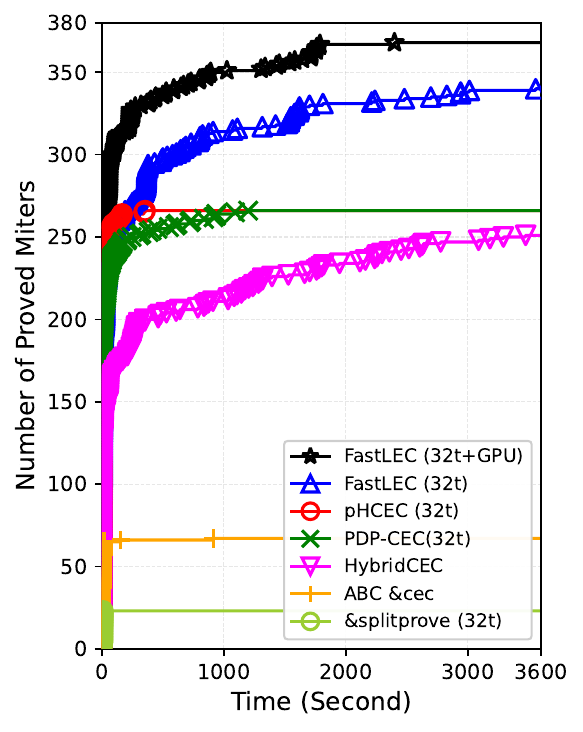}

        \caption{Comparing with SOTA.}\label{fig:cdf-sota}
    \end{minipage}\hfill    
    \begin{minipage}{0.48\textwidth}
        \centering
        \includegraphics[width=0.55\linewidth]{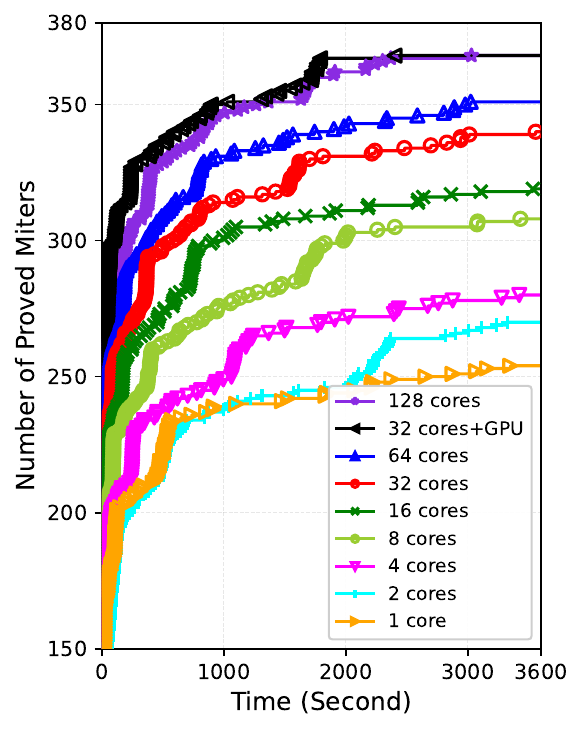}
        \caption{With different resources.}\label{fig:cdf-spd}
    \end{minipage}

\end{figure}

To convey overall trends without listing per-instance results, 
Fig.~\ref{fig:cdf-sota} presents the cumulative distribution function (CDF) 
of solved miters, where steeper curves indicate higher efficiency. 
Hybrid-engine approaches (\hybridcec, \dac, \phcec, and \fastLEC) solve many 
benchmarks quickly, but their progress slows on harder cases when strategy 
choices or engine capabilities become limiting. Overall, \fastLEC achieves the 
best combined proving power and scalability among all evaluated solvers.
The knee point around 250 seconds marks the transition from structurally easy equivalence pairs to deep arithmetic cones, where solver performance begins to diverge across engines.

Moreover, with the help of a single GPU, \fastLEC(32t+GPU) achieves the strongest performance across all difficulty levels: it maintains full coverage on Easy and Medium cases while reducing PAR2 by roughly 72.8\%, and increases solved Hard instances from 93 to a full 121. Overall, these results show that GPU-accelerated ES effectively cuts deep search costs and substantially strengthens the solver’s total proving capacity.


\vspace{-0.5cm}
\subsection{Speedup Analysis for \fastLEC}

Speedup is an important metric for evaluating the scalability of a parallel algorithm.
The \textit{$t$-thread speedup} for a parallel algorithm is defined as the single-core runtime $\eta(1)$ divided by the $t$ multi-core runtime $\eta(t)$.

Table~\ref{tbl:speedup} presents the performance of \fastLEC across various core configurations, including 1, 2, 4, 8, 16, 32, 64, and 128 cores.
Additionally, we provide the maximum (\texttt{Max}), minimum (\texttt{Min}), and geometric mean (\texttt{Geo}) of the speedup for each category.  Noting that the significant improvement in the 2-threading speedup column, which can be attributed to the introduction of the portfolio approach. This strategy effectively assigns hard-to-prove pairs to the most suitable engine, leading to enhanced performance.
Thus, we should focus more on the PAR2 and Geo metric.

From the CDF plot in Fig.\ref{fig:cdf-spd}, we observe that \fastLEC demonstrates strong speedup performance. As complexity increases, the Par2 score decreases more  significantly. Meanwhile, \fastLEC exhibits substantial enhancement with the deployment of more than 8 CPU cores.

\begin{table}
\centering
\setlength{\tabcolsep}{15pt}

\caption{Performance of \fastLEC under various CPU cores\label{tbl:speedup}}
\scalebox{0.55}{
\renewcommand{\arraystretch}{0.75}
\begin{tabular}{cccccccccc}

\toprule
Class & Metric & 1t & 2t & 4t & 8t & 16t & 32t & 64t & 128t \\
\midrule
\multirow{5}{*}{\makecell{Easy\\(127)}}
 & \# Proved & 127 & 127 & 127 & 127 & 127 & 127 & 127 & 127 \\
 & PAR2 & 14.47 & 27.40 & 15.15 & 14.52 & 13.89 & 14.40 & 13.57 & 13.00 \\ \cmidrule(r){2-10}
 & Max & 1 & 1.15 & 1.27 & 1.26 & 1.37 & 1.44 & 1.61 & 1.59 \\
 & Min & 1 & 0.03 & 0.30 & 0.33 & 0.36 & 0.42 & 0.48 & 0.52 \\
 & Geo & 1.0 & 0.21 & 0.46 & 0.47 & 0.53 & 0.58 & 0.69 & 0.80 \\
\midrule
\multirow{5}{*}{\makecell{Medium\\(120)}}
 & \# Proved & 88 & 104 & 108 & 120 & 120 & 120 & 120 & 120 \\
 & PAR2 & 2042.67 & 1370.06 & 992.19 & 270.87 & 123.05 & 62.11 & 33.43 & 18.55 \\ \cmidrule(r){2-10}
 & Max & 1 & 1.42 & 2.31 & 6.43 & 13.69 & 25.60 & 45.66 & 80.87 \\
 & Min & 1 & 0.07 & 0.58 & 0.64 & 0.74 & 0.87 & 1.12 & 1.16 \\
 & Geo & 1.0 & 0.59 & 1.33 & 2.46 & 3.79 & 5.14 & 6.87 & 8.97 \\
\midrule
\multirow{5}{*}{\makecell{\textbf{Hard}\\(121)}}
 & \# Proved & 39 & 39 & 45 & 61 & 72 & 93 & 104 & 121 \\
 & PAR2 & 5208.29 & 5235.80 & 4838.12 & 3972.17 & 3338.35 & 2241.89 & 1551.44 & 537.12 \\ \cmidrule(r){2-10}
 & Max & 1 & 1.08 & 2.11 & 6.22 & 12.58 & 24.01 & 46.31 & 85.80 \\
 & Min & 1 & 0.21 & 1.12 & 1.28 & 1.32 & 1.38 & 1.47 & 1.44 \\
 & Geo & 1.0 & 0.83 & 1.68 & 3.27 & 5.13 & 7.16 & 8.97 & 11.57 \\
\midrule
\multirow{5}{*}{\makecell{ALL\\(368)}}
 & \# Proved & 254 & 270 & 280 & 308 & 319 & 340 & 351 & 368 \\
 & PAR2 & 2383.59 & 2177.77 & 1919.56 & 1399.40 & 1142.59 & 762.36 & 525.71 & 187.15 \\ \cmidrule(r){2-10}
 & Max & 1 & 1.42 & 2.31 & 6.43 & 13.69 & 25.60 & 46.31 & 85.80 \\
 & Min & 1 & 0.03 & 0.30 & 0.33 & 0.36 & 0.42 & 0.48 & 0.52 \\
 & Geo & 1.0 & 0.37 & 0.81 & 1.12 & 1.49 & 1.83 & 2.28 & 2.79 \\
\bottomrule
\hline
\end{tabular}
}
\end{table}

\vspace{-0.65cm}
\subsection{Effectiveness of Engine Scheduling}
\vspace{-0.25cm}

To assess the effectiveness of our scheduling engine, we evaluate two \fastLEC variants: 
\textbf{$V_1$} (only selection, no scheduling), which replaces the scheduling policy with the heuristic used in \hybridcec/\phcec/\dac, and 
\textbf{$V_2$} (poor scheduling strategy), a basic portfolio that splits threads evenly between ES and SAT and shares 1 thread from SAT to BDD. The 32-thread results are summarized in Table~\ref{tbl:variants}.
\#P denotes \#Proved; $\Delta$ denotes the change.

The scheduling mechanism significantly improves performance, particularly on harder instances, by selecting the appropriate engine and distributing resources effectively. 
On easy cases, however, its overhead becomes noticeable, and $V_1$ achieves lower PAR2. 
For easy and medium classes, the portfolio approach ($V_2$) performs well, even better than \fastLEC, because SAT, BDD, and ES offer strong complementarity, and engine choice dominates runtime. 
For hard instances, both engine selection and resource allocation are critical, and the scheduling method enables \fastLEC to outperform both variants.

\begin{table}[t]
\centering
\caption{Performance of the two variants of \fastLEC.\label{tbl:variants}}
\setlength{\tabcolsep}{4pt}

\scalebox{0.6}{
\renewcommand{\arraystretch}{0.8}

\begin{tabular}{rlrlrlrlrlr}
\toprule
\multirow{2}{*}{Prover} &
\multicolumn{2}{c}{Easy (127)} &
\multicolumn{2}{c}{Medium (120)} &
\multicolumn{2}{c}{Hard (121)} &
\multicolumn{2}{c}{All (368)} \\
\cmidrule(r){2-9}
& \#P ($\Delta$) & PAR2 ($\Delta$) 
& \#P ($\Delta$) & PAR2 ($\Delta$)
& \#P ($\Delta$) & PAR2 ($\Delta$)
& \#P ($\Delta$) & PAR2 ($\Delta$) \\
\midrule

\fastLEC(32t) 
& 127  & 14.40 
& 120  & 62.11
& 93   & 2241.89
& 340  & 762.36 \\

$V_1$(32t)    
& 127 (0) & 5.35 (-62.85\%) 
& 88 (-32) & 1927.55 (3003.68\%)
& 52 (-41) & 4179.91 (86.45\%)
& 267 (-73) & 2004.77 (162.97\%) \\

$V_2$(32t)    
& 127 (0) & 15.33 (6.45\%)
& 120 (0) & 120.63 (94.24\%)
& 81 (-12) & 2999.52 (33.79\%)
& 328 (-12) & 1030.88 (35.22\%) \\

\bottomrule
\end{tabular}
}
\end{table}

\vspace{-0.65cm}
\subsection{Effectiveness of Structure-Aware SAT Partitioning}

\vspace{-0.25cm}


To obtain a clean measurement of SAT scalability, ES and BDD are disabled, and only the
SAT engine is evaluated under the structure aware partitioning framework. We select ten
XOR intensive sub-mitters that $\hat{t}_{\mathsf{SAT}} < \hat{t}_{\mathsf{BDD}}$ and the 1-thread runtime for SAT $t^{1t}_{SAT} > 500 s$ to avoid confounding influence. Table \ref{tbl:SP} reports the PAR2 scores averaged over thread
counts from 1 to 128, revealing the intrinsic asymptotic behavior of the SAT component and
confirming its monotonic scalability.

\begin{table}[t]
\caption{Performance Analysis for the Structure-Aware Partitioning SAT Solver\label{tbl:SP}}
\centering
\scalebox{0.55}{
\setlength{\tabcolsep}{3pt}
\renewcommand{\arraystretch}{0.8}

\begin{tabular}{ccrrrrrrrrr}
\toprule
ID & num\_vars & 1t & 2t & 4t & 8t & 16t & 32t & 64t & 128t \\
\midrule
1 & 1349 & 632.28 & 415 & 227.49 & 116.56 & 60.16 & 50.75 & 33.9 & 18.84 \\
2 & 1359 & 593.96 & 446.78 & 234.96 & 123.55 & 62.09 & 73.28 & 28.68 & 17.64 \\
3 & 1364 & 534.04 & 303.7 & 181.73 & 106.84 & 55.68 & 64.06 & 34.97 & 21.94 \\
4 & 1594 & -- & 3188.45 & 1630.19 & 792.21 & 480.34 & 310.77 & 188.99 & 87.86 \\
5 & 1603 & -- & -- & 1449.07 & 924.43 & 582.93 & 343.54 & 199.89 & 131.81 \\
6 & 1603 & -- & 2856.8 & 1358.31 & 892.7 & 605.84 & 385.9 & 143.43 & 101.94 \\
7 & 1842 & -- & -- & -- & -- & -- & -- & 1130.47 & 567.04 \\
8 & 2287 & 2278.79 & 1732.17 & 999.92 & 464.62 & 212.59 & 146.11 & 94.27 & 31.18 \\
9 & 2973 & -- & -- & -- & -- & -- & -- & 2108.05 & 1394.81 \\
10 & 3005 & -- & 3152.65 & 1775.62 & 698.1 & 400.07 & 219.2 & 128.74 & 101.73 \\
\hline
\multicolumn{2}{c}{PAR2} & 4723.91&	3369.55&	2225.73	&1851.9&	1685.97&	1599.36&	409.14&	247.48 \\
\multicolumn{2}{c}{Speedup} & -- & 1.40x &	2.12x 	& 2.55x  & 	2.80x &	2.95x &	11.55x  &	19.09x &\\

\bottomrule
\end{tabular}
}
\end{table}

\vspace{-0.5cm}
\subsection{Feature Contribution to the Prediction}
\vspace{-0.25cm}

\begin{figure}[t]
    \centering
    \includegraphics[width=0.4\linewidth]{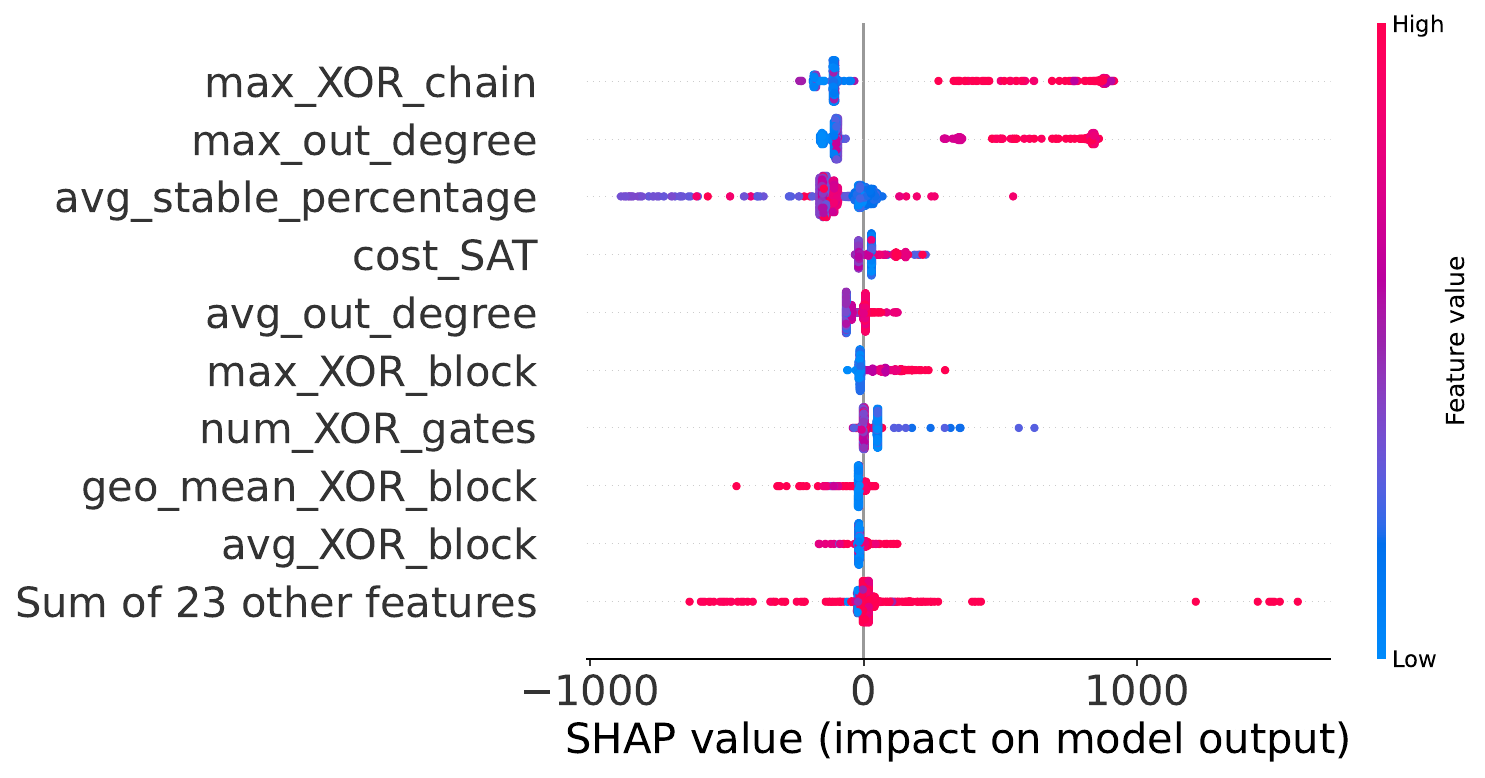}
    \includegraphics[width=0.4\linewidth]{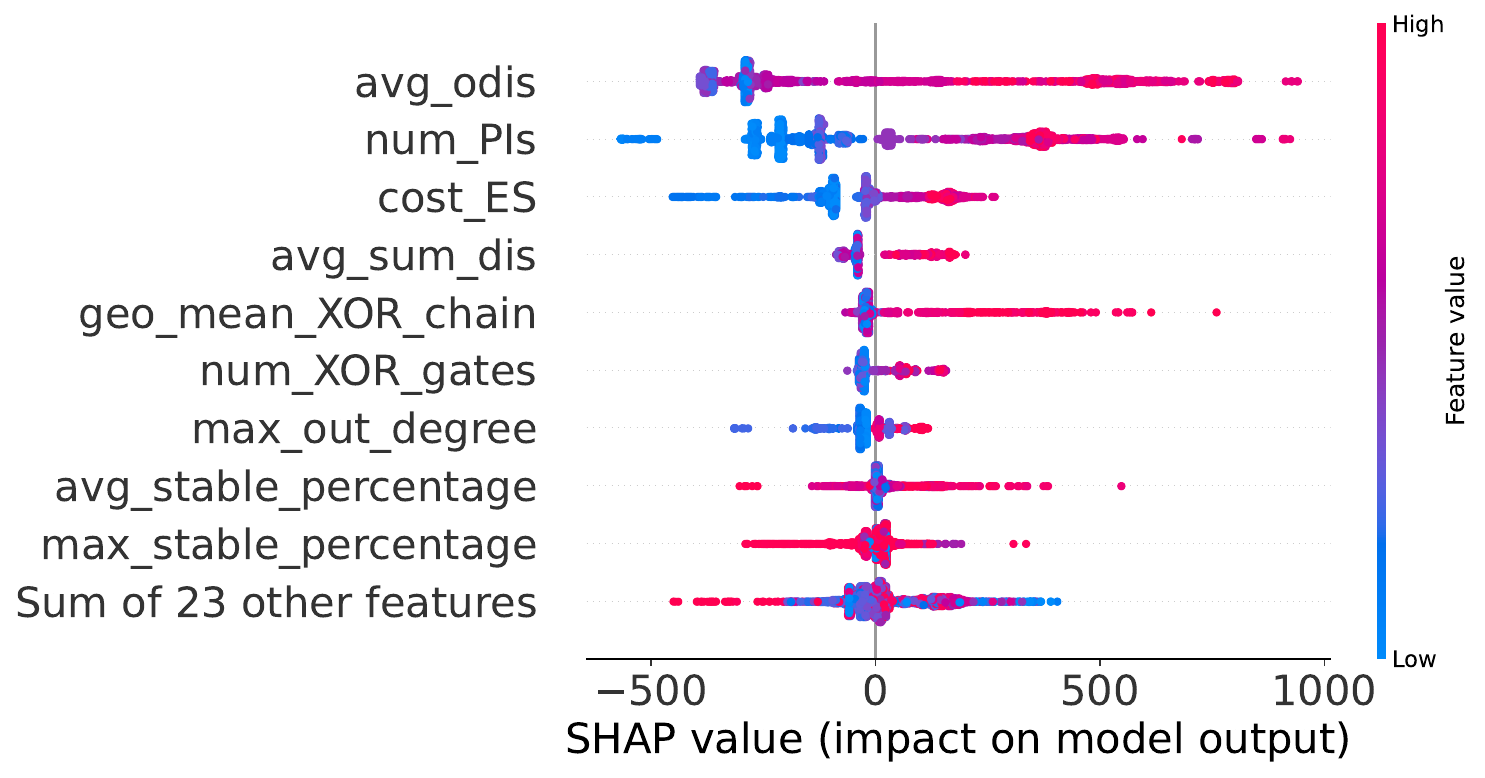}
    \caption{
    The influence of input features on the regression model. Each dot shows a feature’s SHAP value for one sample, colored by its original value. Positive SHAP values indicate contributions that increase the predicted runtime.
    }
    \label{fig:shap}
\end{figure}

We present in Fig.~\ref{fig:shap} each feature's marginal contributions to the models' prediction using the SHAP package~\cite{lundberg2020local}, which explains the output of machine learning models based on the classic Shapley values~\cite{shapley1953value} from game theory. Shapley values fairly attribute credit to individual features for their contributions to the model's overall prediction. Computing exact Shapley values can be computationally intensive. We refer readers to the previous work~\cite{lundberg2020local} for detailed definitions and implementations of Shapley values. Fig.~\ref{fig:shap} presents the SHAP values of the top 9 features that most influence the prediction of the regression model, highlighting the most important properties of miters. 


The prediction models for SAT and BDD performance share only three common features as shown in Fig.~\ref{fig:shap}, including \texttt{max\_out\_degree}, \texttt{num\_XOR\_gates}, and \texttt{avg\_stable\_percentage}. 
For SAT, \texttt{max\_XOR\_chain}, \texttt{max\_out\_degree}, and \texttt{avg\_stable\_percentage} emerge as the most influential predictors. This is consistent with the operational principles of CDCL solving, where shorter XOR dependencies and easier structural elements facilitate localized propagation and effective conflict-driven learning, thereby reducing search divergence.

For BDDs, the leading contributors are \texttt{avg\_odis}, \texttt{num\_PIs}, and \texttt{cost\_ES}. These features characterize the global connectivity that drives decision diagram growth. Higher \texttt{avg\_odis} reflects stronger inter-node interactions that enlarge intermediate BDDs, while \texttt{num\_PIs} determines the dimensionality of the decision space and \texttt{cost\_ES} serves as a proxy for the underlying structural complexity. Unlike SAT, whose behavior is shaped largely by local conflict reasoning, BDD performance is driven by global structural dependencies, often leading to significant variation in diagram size and runtime.

\vspace{-0.45cm}
\section{Conclusion}
\vspace{-0.35cm}

We presented \fastLEC, a learned, hybrid CEC framework that coordinates SAT, BDD, and CPU/GPU ES through predictive scheduling and datapath-aware decomposition. The resulting prover achieves substantial gains over existing sequential and parallel tools, with marked improvements in scalability and coverage. Looking forward, we plan to extend \fastLEC toward distributed, multi-node verification, enabling heterogeneous engines to cooperate at cluster scale for even larger industrial datapath designs.

\newpage
%
%
%
\bibliographystyle{splncs04}
\bibliography{cec}

@inproceedings{mishchenko2008scalable,
  title={Scalable and scalably-verifiable sequential synthesis},
  author={Mishchenko, Alan and Case, Michael and Brayton, Robert and Jang, Stephen},
  booktitle={2008 IEEE/ACM International Conference on Computer-Aided Design},
  pages={234--241},
  year={2008},
  organization={IEEE}
}

@inproceedings{vasicek2017relaxed,
  title={Relaxed equivalence checking: a new challenge in logic synthesis},
  author={Vasicek, Zdenek},
  booktitle={2017 IEEE 20th International Symposium on Design and Diagnostics of Electronic Circuits \& Systems (DDECS)},
  pages={1--6},
  year={2017},
  organization={IEEE}
}

@article{jarratt2011engineering,
  title={Engineering change: an overview and perspective on the literature},
  author={Jarratt, TAW and Eckert, Claudia M and Caldwell, Nicholas HM and Clarkson, P John},
  journal={Research in engineering design},
  volume={22},
  pages={103--124},
  year={2011},
  publisher={Springer}
}

@inproceedings{agrawal1996characteristic,
  title={Characteristic polynomial method for verification and test of combinational circuits},
  author={Agrawal, Vishwani D and Lee, David},
  booktitle={Proceedings of 9th International Conference on VLSI Design},
  pages={341--342},
  year={1996},
  organization={IEEE}
}

@article{kuehlmann2002robust,
  title={Robust Boolean reasoning for equivalence checking and functional property verification},
  author={Kuehlmann, Andreas and Paruthi, Viresh and Krohm, Florian and Ganai, Malay K},
  journal={IEEE Transactions on Computer-Aided Design of Integrated Circuits and Systems},
  volume={21},
  number={12},
  pages={1377--1394},
  year={2002},
  publisher={IEEE}
}

@article{tseitin1983complexity,
  title={On the complexity of derivation in propositional calculus},
  author={Tseitin, Grigori S},
  journal={Automation of reasoning: 2: Classical papers on computational logic 1967--1970},
  pages={466--483},
  year={1983},
  publisher={Springer}
}

@inproceedings{kuehlmann2004dynamic,
  title={Dynamic transition relation simplification for bounded property checking},
  author={Kuehlmann, Andreas},
  booktitle={IEEE/ACM International Conference on Computer Aided Design, 2004. ICCAD-2004.},
  pages={50--57},
  year={2004},
  organization={IEEE}
}

@inproceedings{lv2012formal,
  title={Formal verification of Galois field multipliers using computer algebra techniques},
  author={Lv, Jinpeng and Kalla, Priyank},
  booktitle={2012 25th International Conference on VLSI Design},
  pages={388--393},
  year={2012},
  organization={IEEE}
}

@article{mishchenko2007abc,
  title={ABC: A system for sequential synthesis and verification},
  author={Mishchenko, Alan and others},
  journal={URL http://www. eecs. berkeley. edu/alanmi/abc},
  volume={17},
  year={2007}
}

@inproceedings{chen2023integrating,
  title={Integrating exact simulation into sweeping for datapath combinational equivalence checking},
  author={Chen, Zhihan and Zhang, Xindi and Qian, Yuhang and Xu, Qiang and Cai, Shaowei},
  booktitle={2023 IEEE/ACM International Conference on Computer Aided Design (ICCAD)},
  pages={1--9},
  year={2023},
  organization={IEEE}
}

@inproceedings{wu2006potential,
  title={The Potential and Limitation of Probability-Based Combinational Equivalence Checking},
  author={Wu, Shih-Chieh and Wang, Chun-Yao and Hsieh, Jan-An},
  booktitle={2006 15th Asian Test Symposium},
  pages={103--108},
  year={2006},
  organization={IEEE}
}

@article{wu2008novel,
  title={Novel probabilistic combinational equivalence checking},
  author={Wu, Shih-Chieh and Wang, Chun-Yao and Chen, Yung-Chih},
  journal={IEEE transactions on very large scale integration (VLSI) systems},
  volume={16},
  number={4},
  pages={365--375},
  year={2008},
  publisher={IEEE}
}

@inproceedings{bjesse2004dag,
  title={DAG-aware circuit compression for formal verification},
  author={Bjesse, Per and Boralv, Arne},
  booktitle={IEEE/ACM International Conference on Computer Aided Design, 2004. ICCAD-2004.},
  pages={42--49},
  year={2004},
  organization={IEEE}
}

@techreport{mishchenko2005fraigs,
  title={FRAIGs: A unifying representation for logic synthesis and verification},
  author={Mishchenko, Alan and Chatterjee, Satrajit and Jiang, Roland and Brayton, Robert K},
  year={2005},
  institution={ERL Technical Report}
}

@article{long2013lec,
  title={LEC: Learning-Driven Data-path Equivalence Checking},
  author={Long, Jiang and Brayton, Robert K and Case, Michael},
  journal={Program Proceedings},
  pages={9},
  year={2013}
}

@article{possani2019parallel,
  title={Parallel combinational equivalence checking},
  author={Possani, Vinicius N and Mishchenko, Alan and Ribas, Renato P and Reis, Andre I},
  journal={IEEE Transactions on Computer-Aided Design of Integrated Circuits and Systems},
  volume={39},
  number={10},
  pages={3081--3092},
  year={2019},
  publisher={IEEE}
}

@inproceedings{DBLP:conf/cav/ZhaoCQ24,
  author       = {Mengyu Zhao and
                  Shaowei Cai and
                  Yuhang Qian},
  editor       = {Arie Gurfinkel and
                  Vijay Ganesh},
  title        = {Distributed {SMT} Solving Based on Dynamic Variable-Level Partitioning},
  booktitle    = {Computer Aided Verification - 36th International Conference, {CAV}
                  2024, Montreal, QC, Canada, July 24-27, 2024, Proceedings, Part {I}},
  series       = {Lecture Notes in Computer Science},
  volume       = {14681},
  pages        = {68--88},
  publisher    = {Springer},
  year         = {2024},
  url          = {https://doi.org/10.1007/978-3-031-65627-9\_4},
  doi          = {10.1007/978-3-031-65627-9\_4},
  bibsource    = {dblp computer science bibliography, https://dblp.org}
}

@inproceedings{heule2016solving,
  title={Solving and verifying the boolean pythagorean triples problem via cube-and-conquer},
  author={Heule, Marijn JH and Kullmann, Oliver and Marek, Victor W},
  booktitle={International Conference on Theory and Applications of Satisfiability Testing},
  pages={228--245},
  year={2016},
  organization={Springer}
}

@inproceedings{heule2011cube,
  title={Cube and conquer: Guiding CDCL SAT solvers by lookaheads},
  author={Heule, Marijn JH and Kullmann, Oliver and Wieringa, Siert and Biere, Armin},
  booktitle={Haifa Verification Conference},
  pages={50--65},
  year={2011},
  organization={Springer}
}

@article{schreiber2024mallobsat,
  title={MallobSat:: Scalable SAT Solving by Clause Sharing},
  author={Schreiber, Dominik and Sanders, Peter},
  journal={Journal of Artificial Intelligence Research},
  volume={80},
  year={2024},
  publisher={AI Access Foundation}
}

@article{zhang2022parkissat,
  title={Parkissat: Random shuffle based and pre-processing extended parallel solvers with clause sharing},
  author={Zhang, Xindi and Chen, Zhihan and Cai, Shaowei},
  journal={SAT COMPETITION 2022},
  pages={51},
  year={2022}
}

@article{karypis1997metis,
  title={METIS: A software package for partitioning unstructured graphs, partitioning meshes, and computing fill-reducing orderings of sparse matrices},
  author={Karypis, George and Kumar, Vipin},
  year={1997}
}

@inproceedings{chatterjee2010equipe,
  title={EQUIPE: Parallel equivalence checking with GP-GPUs},
  author={Chatterjee, Debapriya and Bertacco, Valeria},
  booktitle={2010 IEEE International Conference on Computer Design},
  pages={486--493},
  year={2010},
  organization={IEEE}
}

@article{chowdhary1999extraction,
  title={Extraction of functional regularity in datapath circuits},
  author={Chowdhary, Amit and Kale, Sudhakar and Saripella, Phani K and Sehgal, Naresh K and Gupta, Rajesh K},
  journal={IEEE Transactions on Computer-Aided Design of Integrated Circuits and Systems},
  volume={18},
  number={9},
  pages={1279--1296},
  year={1999},
  publisher={IEEE}
}

@article{kamaraju2010power,
  title={Power optimized ALU for efficient datapath},
  author={Kamaraju, Marellasv and Kishore, K Lal and Tilak, AVN},
  journal={International Journal of Computer Applications},
  volume={11},
  number={11},
  pages={39--43},
  year={2010},
  publisher={Citeseer}
}

@inproceedings{zhou25dac,
  title={DPCEC: A Dynamic Partitioning Combinational Equivalence Checking Parallel Engine},
author={Zhou, Shuai and Zhang, Weikang and Zhang, Xindi and Jiang, Zite and Cai, Shaowei and You, Haihang},
  booktitle={2025 62th ACM/IEEE Design Automation Conference (DAC)},
  pages={1--6},
  year={2025},
  organization={IEEE}
}

@article{chen2024datapath,
  title={Datapath Combinational Equivalence Checking With Hybrid Sweeping Engines and Parallelization},
  author={Chen, Zhihan and Zhang, Xindi and Qian, Yuhang and Cai, Shaowei},
  journal={arXiv preprint arXiv:2501.14740},
  year={2024}
}

@inproceedings{chen2016xgboost,
  title={Xgboost: A scalable tree boosting system},
  author={Chen, Tianqi and Guestrin, Carlos},
  booktitle={Proceedings of the 22nd acm sigkdd international conference on knowledge discovery and data mining},
  pages={785--794},
  year={2016}
}

@article{lundberg2020local,
  title={From local explanations to global understanding with explainable AI for trees},
  author={Lundberg, Scott M and Erion, Gabriel and Chen, Hugh and DeGrave, Alex and Prutkin, Jordan M and Nair, Bala and Katz, Ronit and Himmelfarb, Jonathan and Bansal, Nisha and Lee, Su-In},
  journal={Nature machine intelligence},
  volume={2},
  number={1},
  pages={56--67},
  year={2020},
  publisher={Nature Publishing Group}
}

@article{shapley1953value,
  title={A value for n-person games},
  author={Shapley, Lloyd S and others},
  year={1953},
  publisher={Princeton University Press Princeton},
}

@article{somenzi1998cudd,
  title={CUDD: CU decision diagram package release 2.3. 0},
  author={Somenzi, Fabio},
  journal={University of Colorado at Boulder},
  volume={621},
  year={1998}
}

@article{JMLR:v23:21-0888,
  author  = {Marius Lindauer and Katharina Eggensperger and Matthias Feurer and André Biedenkapp and Difan Deng and Carolin Benjamins and Tim Ruhkopf and René Sass and Frank Hutter},
  title   = {SMAC3: A Versatile Bayesian Optimization Package for Hyperparameter Optimization},
  journal = {Journal of Machine Learning Research},
  year    = {2022},
  volume  = {23},
  number  = {54},
  pages   = {1--9},
  url     = {http://jmlr.org/papers/v23/21-0888.html}
}

@inproceedings{van2015sylvan,
  title={Sylvan: Multi-core decision diagrams},
  author={Van Dijk, Tom and Van De Pol, Jaco},
  booktitle={International Conference on Tools and Algorithms for the Construction and Analysis of Systems},
  pages={677--691},
  year={2015},
  organization={Springer}
}

@article{chen2025datapath,
  title={Datapath combinational equivalence checking with hybrid sweeping engines and parallelization},
  author={Chen, Zhihan and Zhang, Xindi and Qian, Yuhang and Cai, Shaowei},
  journal={ACM Transactions on Design Automation of Electronic Systems},
  volume={31},
  number={1},
  pages={1--27},
  year={2025},
  publisher={ACM New York, NY}
}

@inproceedings{seger2021formal,
  title={Formal verification of complex data paths: An industrial experience},
  author={Seger, Carl-Johan H},
  booktitle={International Symposium on Formal Methods},
  pages={697--716},
  year={2021},
  organization={Springer}
}

@article{kaufmann2020incremental,
  title={Incremental column-wise verification of arithmetic circuits using computer algebra},
  author={Kaufmann, Daniela and Biere, Armin and Kauers, Manuel},
  journal={Formal Methods in System Design},
  volume={56},
  number={1},
  pages={22--54},
  year={2020},
  publisher={Springer}
}
%




\end{document}